\documentclass[preprint,nocopyrightspace]{sigplanconf}

\usepackage{amsmath}
\usepackage{amssymb}
\usepackage{listings}

\lstdefinelanguage{ML}{
  alsoletter={*},
  morekeywords={datatype, of, if, *},
  sensitive=true,
  morecomment=[s]{/*}{*/},
  morestring=[b]"
}

\lstdefinelanguage{scala}{
  alsoletter={@,=,>},
  morekeywords={abstract, case, catch, class, def, do, else, extends, false, final, finally, for, if, implicit, import, match, new, null, object, 
override, package, private, protected, requires, return, sealed, super, this, throw, trait, try, true, type, val, var, while, with, yield, domain, 
postcondition, precondition,invariant, constraint, assert, forAll,  _, return, @generator, ensure, require, ensuring,=>,
certainly, possibly, certify, errorBound, assertBound, jacobian, derivative},
  sensitive=true,
  morecomment=[l]{//},
  morecomment=[s]{/*}{*/},
  showstringspaces=false,
  columns=fullflexible,
  mathescape=true,
  numberstyle=\tiny,
  basicstyle=\codestyle,
  numbersep=5pt,
  stepnumber=2,
  numbers=left,                   
  morestring=[b]"
}

\newcommand{\codestyle}{\small}

\newcommand{\pseudostyle}{\sffamily}

\lstdefinelanguage{pseudo}{
  alsoletter={@,=,>},
  morekeywords={abstract, case, catch, class, def, do, else, extends, false, final, finally, for, if, implicit, import, match, new, null, object, 
override, package, private, protected, requires, return, sealed, super, this, throw, trait, try, true, type, val, var, while, yield, domain, 
postcondition, invariant, constraint, assert, forAll,  _, return, @generator, ensure, require, ensuring,=>, 
certainly, possibly, certify, errorBound, assertBound, jacobian, derivative, Input, Output},
  sensitive=true,
  morecomment=[l]{//},
  morecomment=[s]{/*}{*/},
  showstringspaces=false,
  columns=fullflexible,
  mathescape=true,
  numberstyle=\tiny,
  basicstyle=\pseudostyle,
  numbers=left,                   
  morestring=[b]"
}

\usepackage{color}
\usepackage{multirow}
\usepackage{graphicx}

\usepackage[normalem]{ulem} 
\newcommand{\code}[1] {\lstinline!#1!}

\begin{document}



\title{On Sound Compilation of Reals}

\authorinfo{Eva Darulova}
           {EPFL}
            {eva.darulova@epfl.ch}
 \authorinfo{Viktor Kuncak}
            {EPFL}
            {viktor.kuncak@epfl.ch}
\vspace{-20pt}
\maketitle

\begin{abstract}
Writing accurate numerical software is hard because of many
sources of unavoidable uncertainties, including finite
numerical precision of implementations.  We present a
programming model where the user writes a program in a
real-valued implementation and specification language that
explicitly includes different types of uncertainties.  We
then present a compilation algorithm that generates a
conventional implementation that is guaranteed to meet the
desired precision with respect to real numbers.  Our
verification step generates verification conditions that
treat different uncertainties in a unified way and encode
reasoning about floating-point roundoff errors into
reasoning about real numbers. Such verification conditions
can be used as a standardized format for verifying the
precision and the correctness of numerical programs.  Due to
their often non-linear nature, precise reasoning about such
verification conditions remains difficult. We show that
current state-of-the art SMT solvers do not scale well to
solving such verification conditions.  We propose a new
procedure that combines exact SMT solving over reals with
approximate and sound affine and interval arithmetic. We
show that this approach overcomes scalability limitations of
SMT solvers while providing improved precision over affine
and interval arithmetic.  Using our initial implementation
we show the usefullness and effectiveness of our approach on
several examples, including those containing non-linear
computation.
\end{abstract}



\section{Introduction}

Writing numerical programs is difficult, in part because 
the programmer needs to deal not only with the correctness
of the algorithm but also with unavoidable uncertainties. Program inputs may not be exactly known
because they come from physical experiments or were measured by an embedded sensor. The
computation itself suffers from roundoff errors at each step, because of the use of
finite-precision arithmetic. In addition, resources like energy
may be scarce so that only a certain number of bits are available for the numerical datatype.
At the same time, 
the computed results can have far-reaching consequences if used to control, for example,
a vehicle or a nuclear power plant.
It is therefore of great importance to
improve confidence in numerical code \cite{leroy11verifSquare}. One of the first
challenges in doing this is that most of our automated 
reasoning tools work with real arithmetic, whereas the code is implemented in
finite-precision (typically, floating
point) arithmetic.
Many current approaches to verifying numerical programs start with the floating-point
implementation and then try to verify the absense of (runtime) errors. However,
the absence of errors by itself does not guarantee that program behavior matches
the desired specification expressed using real numbers. Fundamentally, the source code
semantics is expressed in terms of data types such as floating points. This is further
problematic for compiler optimizations, because, e.g.,
the associativity law is unsound with respect to such source code semantics.

In this paper we advocate a natural but ambitious alternative:
source code programs should be expressed in terms of ideal, mathematical real numbers.
In our system, the programmer writes a program using a \code{Real} data type and states
the desired postconditions, then specifies
uncertainties explicitly in pre- and postconditions as well as the desired target precision on the return
values.
It is then up to our trustworthy compiler to check, taking into account all uncertainties
and their propagation, that the desired precision can be soundly realized in a finite precision
implemenation. If so, the compiler chooses and emits one such finite-precision implementation.

Following this model, we derive verification conditions that encode reasoning about
floating-point roundoff errors into reasoning about real numbers. Our verification conditions
explicitly separate the {\em ideal} program without external uncertainties and roundoffs
from the {\em actual} program, which is actually executed in finite precision with possibly
noisy inputs. Additionally, our constraints are fully parametric in the floating-point
precision and can thus be used with different floating-point hardware configurations,
as well as used for determining the minimum precision needed.

To summarize, the view of source code as functions over real numbers has several advantages:
\begin{itemize}
  \item Programmers can, for the most part, reason in real arithmetic and not floating-point arithmetic.
   We thus achieve separation of the design of algorithms (which may still be approximate) from
   their realization using finite precision computations.

  \item We can verify the ideal meaning of programs using techniques developed to reason over
    real numbers, which are more scalable than techniques that directly deal with floating
    point arithmetic.

  \item The approach allows us to quantify the deviation of implementation outputs from ideal ones, instead
  of merely proving e.g. range bounds of floating-point variables which is used in simpler static analyses.

  \item The compiler for reals is free to do optimizations as long as they preserve the precision
  requirements. This allows the compiler to apply, for example, associativity of arithmetic, or
  even select different approximation schemes for transcendental functions.

  \item In addition to roundoff errors, the approach also allows the developer to quantify
    program behavior in the face of external uncertainties such as input measurement errors.
\end{itemize}

Using our verification condition generation approach, the
correctness and the precision of compilation for small
programs can be directly verified using an SMT solver such
as Z3~\cite{De-Moura2008} (see Section~\ref{sec:bsplines});
this capability will likely continue to improve as the
solvers advance.  However, the complexity of the generated
verification conditions for larger programs is still out of
reach of such solvers and we believe that specialized
techniques are and will continue to be necessary for this
task. This paper presents two specialized techniques that
improve the feasibility of the verification task.  The first
technique performs local approximation and is effective even
in benchmarks containing nonlinear arithmetic; the second
technique specifically handles conditional expressions.

\subsection{Solving Non-Linear Constraints}

Nonlinear arithmetic poses a significant challenge for
verification because it cannot directly be handled using
Simplex-like algorithms embedded inside SMT
solvers. Although interesting relevant fragments are
decidable and are supported by solvers, their complexity is
much higher. Unfortunately, non-linear arithmetic is
ubiquitous in numerical software.  Furthermore, our
verification conditions add roundoff errors to arithmetic
expressions, so the resulting constraints grow further in
complexity, often becoming out of reach of solvers. An
alternative to encoding into SMT solver input is to use a
sound and overapproximating arithmetic model such as
interval or affine arithmetic~\cite{stolfi97}. However, when
used by itself on non-linear code, these approaches yield
too pessimistic results to be useful.

We show that we can combine range arithmetic computation
with SMT solving to overcome the limitations of each of the
individual techniques when applied to non-linear arithmetic.
We obtain a sound, precise, and somewhat more scalable
procedure.  During range computation, our technique also
checks for common problems such as overflow, division by
zero or square root of a negative number, emitting the
corresponding warnings.  Additionally, because the procedure
is a forward computation, it is suitable for automatically
generating function summaries containing output ranges and
errors of a function. From the point of view of the logical
encoding of the problem, range arithmetic becomes a specialized
method to perform approximate quantifier elimination of bounded
variables that describe the uncertainty.

\subsection{Sound Compilation of Conditional Branches}

In the presence of uncertainties, conditional braches become another verification
challenge. The challenge is that the ideal execution
may follow one branch, but, because of input or roundoff errors, the actual execution follows another. 
This
behaviour may be acceptable, however, if we can show that the error on the output remains within required
bounds. 
Therefore, our approach would directly benefit from automated
analysis of continuity, which was advocated previously
\cite{Westbrook2013}. 
For such continuous functions, our analysis can be done separately for each
program path.
In the absence of knowledge of continuity, we present a new method to check that
different paths taken by 
real-valued and floating point versions of the program still preserves
the desired precision specification. Our check 
does not require continuity (which becomes difficult to prove for non-linear code).
Instead, it directly checks that the difference between the two values on different branches
meets the required precision.
This technique extends our method for handling non-linear arithmetic, so it benefits
from the combination of range arithmetic and SMT solving.

\subsection{Implementation and Evaluation}

We have implemented our compilation and verification
procedure, including the verification condition generation,
analysis of possibly non-linear expressions, and the
handling of conditionals.  Our system is implemented as an
extension of a verifier for functional Scala programs.  The
implementation relies on a range arithmetic implementation
for Scala as well as on the Z3 SMT solver.  

We have evaluated the system on a number of diverse
benchmarks, obtaining promising results.  We are releasing
our benchmarks (and making them available as supplementary
material for reviewers).  We hope that the benchmarks can be
used to compare future tools for error quantification, help
the development of nonlinear solvers, and also present
challenges for more aggressive compilation schemes, with
different number representations and different selection of
numerical algorithms.  To support programming of larger code
fragments our system also supports a modular verification
technique, which handles functions through inlining function
bodies or using their postconditions.  We thus expect that
our technique is applicable to larger code bases as well,
possibly through refactoring code into multiple smaller and
annotated functions.  Even on the benchmarks that we
release, we are aware of no other available system that
would provide the same guarantees with our level of
automation.

\subsection{Summary of Contributions} 

Our overall contribution is an approach for sound compilation of real numbers. Specifically:
\begin{itemize}
 \item We present a real-valued implementation and specification language for numerical programs with uncertainties;
 we define its semantics in terms of verification constraints that they induce.
 We believe that such verification conditions can be used as a standardized format for verifying the precision and the 
 correctness of numerical programs.
 
 \item We develop an approximation procedure for computing precise range and error bounds for nonlinear expressions which
 combines SMT solving with interval arithmetic. We show that such
 an approach significantly improves computed range and error bounds 
 compared to standard interval arithmetic, and scales better than SMT solving alone.
 Our procedure can also be used independently as a more precise alternative to interval arithmetic, and thus
 can perform forward computation without having the desired postconditions.

 \item We 
   describe an approach for soundly computing error bounds in the presence of branches 
   and uncertainties, which ensures soundness of compilation in case the function 
   defined by a program is not known to be continuous.


\item We have implemented our framework and report our
  experience on a set of diverse benchmarks, including
  benchmarks from physics, biology, chemistry, and control
  systems. The results show that our technique is effective
  and that it achieves a synergy of the techniques on which
  it relies.
\end{itemize}

\section{Example}
\label{sec:example}
\begin{figure}
\begin{lstlisting}
def mainFunction(a: Real, b: Real, c: Real): Real = {
    require(4.500005 <= a && a <= 6.5)

    val b = 4.0
    val c = 8.5
    val area = triangleSorted(a, b, c)
    //val area = triangleUnstable(a, b, c)
    area
  } ensuring(res => res +/- 1e-11)

  def triangle(a: Real, b: Real, c: Real): Real = {
    require(a.in(1.0, 9.0) && b.in(1.0, 9.0) && c.in(1.0, 9.0) &&
      a + b > c + 1e-6 && a + c > b + 1e-6 && b + c > a + 1e-6)

    val s = (a + b + c)/2.0
    sqrt(s * (s - a) * (s - b) * (s - c))
  }
  
  def triangleSorted(a: Real, b: Real, c: Real): Real = {
    require(a.in(1.0, 9.0) && b.in(1.0, 9.0) && c.in(1.0, 9.0) && 
      a + b > c + 1e-6 && a + c > b + 1e-6 && b + c > a + 1e-6 &&
      a < c && b < c)

    if (a < b) {
      sqrt((c+(b+a)) * (a-(c-b)) * (a+(c-b)) * (c+(b-a)))/4.0
    } else {
      sqrt((c+(a+b)) * (b-(c-a)) * (b+(c-a)) * (c+(a-b))) / 4.0
    }
  }
\end{lstlisting}
\caption{Computing the area of a triangle.}
\label{fig:example}
\end{figure}

We demonstrate some aspects of our system
on the example in Figure~\ref{fig:example}.
The methods \code{triangle} and \code{triangleSorted} compute the area of a triangle
with side lengths $a, b$ and $c$. The notation \code{a.in(1.0, 9.0)} is short for \code{1.0 < a && a < 9.0}. 
We consider a particular application where the user
may have two side lengths given, and may vary the third. She has two functions available
to do the computation and wants to determine whether either or both satisfy the precision
requirement of 1e-11 on line 9.  Our tool determines that such requirement
needs at least double floating point precision; the challenge then is to establish
that this precision is sufficient to ensure these bounds, given that errors in floating
code accumulate and grow without an \emph{a priori} bound.

Our tool verifies fully automatically that the method \code{triangleSorted} indeed satisfies
the postcondition and generates the source code
with the \code{Double} datatype which also includes a more precise and complete postcondition:

\begin{lstlisting}[mathescape,numbers=none]
    0.01955760939159717 $\le$ res $\wedge$ res $\le$ 12.519984025578283 $\wedge$
      res +/- 8.578997409317759e-12
\end{lstlisting}

To achieve this result, our tool first checks that the precondition of the function
call is satisfied using the Z3 solver. Then, it inlines the body of the function 
\code{triangleSorted} and computes a sound bound on the result's uncertainty with our approximation 
procedure and uses it to show that the postcondition is satisfied. 
The error computation takes into account in a sound way the input uncertainty (here an initial 
roundoff error on the inputs), its propagation and roundoff errors commited at each arithmetic operation.
Additionally, due to the initial roundoff error the comparison on line 24 is not exact, so that
some floating-point computations will take a different branch than their corresponding real-valued
computation. More precisely, the total error when computing the condition is $7.22e-16$, as computed
by our tool. That is, floating-point values that satisfy $a < b + 7.22e-16$ may take the else branch,
even though the corresponding real values would follow the then branch, and similarly in the opposite direction.
Our tool verifies that the difference in the computed result in two branches, due to this divergence between
real arithmetic and floating point arithmetic code, remains within the precision requirement.

 Finally, our tool uses our novel range computation procedure to also compute a more precise output
 range than we could have obtained in interval arithmetic. It then includes this more complete postcondition
 in the generated floating-point code.
 In fact, interval arithmetic alone computes the ranges
$[0.0138, 16.163]$ and $[0.0169, 14.457]$ for using the methods \code{triangle} and \code{triangleSorted}
respectively, seemingly suggesting that two methods perform entirely different computations.
With our technique, the tool computes the same range $[0.0195, 12.52]$ for both
methods, but shows a difference in the absolute error of the computation.
For the method \code{triangle}, the verification fails, as desired, because the computed error ($2.3e-11$) exceeds
the required precision bound. This result is expected---the textbook formula for triangles
is known to suffer from imprecision for flat triangles~\cite{triangle}, which is somewhat rectified in the
method \code{triangleSorted}.

\section{Programs with Reals}
\label{sec:specification}

Each program to be compiled consists of one top-level object with methods written in a
functional subset of the Scala programming language~\cite{Odersky2008}.
All methods are functions over the \code{Real} datatype and the user annotates them 
with pre- and postconditions that explicitly talk about uncertainties.
\code{Real} represents ideal real numbers without any uncertainty. We allow arithmetic
expressions over \code{Real}s with the standard arithmetic operators $\{+, -, *, /, \sqrt{}\}$,
and together with conditionals and function calls they form the body of
methods. Our tool also supports immutable variable declarations as \code{val x = ...}.
This language allows the user to define a computation over real numbers.
Note that this specification language is not executable.

The precondition allows the user to provide a specification of the environment. A complete
environment specification consists of lower and upper bounds for all method parameters and
an upper bound on the uncertainty or noise. Range bounds are expressed with regular comparison
operators. Uncertainty is expressed with the predicate such as \code{x +/- 1e-6}, which denotes that the
variable $x$ is only known up to $1e-6$. Alternatively, the programmer can specify the relative
error as \code{x +/- 1e-7 * x}.
If no noise except for roundoff is present, roundoff errors are automatically added to input variables.

The postcondition can specify the range and the maximum uncertainty accepted on the output.
In addition to the language allowed in the precondition, the postcondition may reference the
errors on inputs directly in the following way: \lstinline$res +/- 3.5 * !x$, which says that the
maximum acceptable error on the output variable $res$ is bounded from above by $3.5$ times the initial
error on $x$.
Whereas the precondition may only talk about the ideal values, the postcondition can also
reference the actual value directly via \code{$\sim$x}. This allows us to assert that runtime
values will not exceed a certain range, for instance.

\paragraph{Floating-point arithmetic} Our tool and technique support in the generated
target code any floating-point precision
and in particular, single and double floating-point precision as defined by the IEEE 754
floating-point standard~\cite{ieee75408}. It is also straightforward to combine it with
techniques to generate fixed-point arithmetic~\cite{Darulova2013}, 
which similarly relies on knowing ranges of variables.
We assume rounding-to-nearest rounding
 mode and that basic arithmetic operations $\lbrace +, -, *, / , \sqrt{} \rbrace$ are rounded
correctly, which means that the result from any such operation must be the
closest representable floating-point number. Hence, provided there is no
overflow, the result of a binary operation $\circ_F$ satisfies
\begin{equation} x \circ_F y = (x \circ_R y)(1 + \delta),   \;\; |\delta| \le
\epsilon_M, \;\; \circ \in \lbrace +, -, *, /  \rbrace \label{eqn:floatabstraction}
\end{equation} where $\circ_R$ is the ideal operation in real numbers and
$\epsilon_M$ is the machine epsilon that determines the upper bound on the
relative error. This model provides a basis for our roundoff error estimates.

\section{Compiling Reals to Finite Precision}
\begin{figure}
\begin{lstlisting}[mathescape,language=pseudo,numbers=none]
Input: spec: specification over Reals, prec: candidate precisions
for fnc $\leftarrow$ spec.fncs
  fnc.vcs = generateVCs(fnc)

spec.fncs.sortBy((f1, f2) => f1 $\subseteq$ f2.fncCalls)  

while prec $\ne \emptyset$ and notProven(spec.fncs)
  precision = prec.nextPrecise
  for fnc $\leftarrow$ spec.fncs
    for vc $\leftarrow$ fnc.vcs
      while vc.hasNextApproximation $\wedge$ notProven(vc)
        approx = getNextApproximation(vc, precision)
        vc.status = checkWithZ3(approx)
  generateSpec(fnc)
generateCode(spec)

Output: floating-point code with generated postconditions
\end{lstlisting}
\caption{Compilation algorithm.}
\label{alg:compilation}
\end{figure}

Given a specification or program over reals and a possible target floating-point precision,
our tool generates code over floating-point numbers that satisfy the given pre- and postconditions.
Figure~\ref{alg:compilation} presents a high-level view of our compilation algorithm.
Our tool first analyses the entire specification and generates one verification condition for
each postcondition to be proven. To obtain a modular algorithm, the tool also generates verification 
conditions that check that at each function call the precondition of the called function is satisfied.
The methods are then sorted by occuring function calls. This allows us to re-use already 
computed postconditions of function calls in a modular analysis.
If the user specified a target precision, the remaining part of the compilation process is perfomed
with respect to this precision, if not or in the case the user specified several possible precisions,
our tool will iteratively select the next more precise precision and attempt to compile the entire program. 
If compilation fails due to unsatisfied assertions,
the next precision is selected. Thus, one task of our algorithm is to automatically determine the necessary
least floating-point precision to ensure the specification is met.
Currently, each precision iteration is computed separately, which is not a big issue
due to a small number alternative floating-point targets. 
We did identify certain shared 
computations between iterations; we can exploit them in the future
for more efficient compilation.
In order for the compilation process to succeed, the specification has to be met with respect to a given
floating-point precision, thus the principal part of our algorithm is spent in verification, which we
describe in Section~\ref{sec:verification}.

We envision that in the future the compilation task will also include automatic precision-preserving 
code optimizations, but in this paper we concentrate on the challenging groundwork of verifying the 
precision of code.

Our tool currently generates Scala code over single, double, double-double 
and quad-double floating-point arithmetic. For double-double and quad-double precision, which were implemented
in software by~\cite{QD2013}, we provide a Scala interface with the generated code.
In case the verification part of compilation fails, we nonetheless generate code (together with a failure report) 
with the best postconditions our tool was able to compute.
The user can then use the generated specifications to gain insight why and where her program does not satify requirements.

Our techniques are not restricted to these floating-point precisions, however, and will work for any
precision that follows the abstraction given in Equation~(\ref{eqn:floatabstraction}).
Furthermore, while we have implemented our tool to accept specifications in a domain specific 
language embedded in Scala and generate code in Scala, all our techniques apply equally to all 
programming languages and hardware that follow the floating-point abstraction we assume (Equation~\ref{eqn:floatabstraction}).

\section{Verifying Real Programs}
\label{sec:verification}
We will now describe the verification part of our compilation algorithm.
In the following we will call the {\em ideal} computation the computation in the absence of
any uncertainties and implemented in a real arithmetic, and the {\em actual} computation
the computation that will finally be executed in finite-precision and with potentially
uncertain inputs.

\subsection{Verification Conditions for Loop-Free Programs}
Each method with its precondition $P$ and postcondition $Q$ implies the following
verification condition:
\begin{equation}
\tag{*}
\label{hornclause}
\forall \vec{x}, \vec{res}, \vec{y}.\; P(\vec{x}) \wedge body(\vec{x}, \vec{y}, \vec{res})
\to Q(\vec{x}, \vec{res})
\end{equation}
where $\vec{x}, \vec{res}, \vec{y}$ denote the input, output and local variables respectively.

\renewcommand{\arraystretch}{1.3}
\begin{table}
  \centering
  \begin{tabular}{ l | l }
    \lstinline$a <= x && x <= b$  &  $x \in [a, b]$ \\

    \lstinline$x +/- k$ &  $x_\circ = x + err_x \wedge err_x \in [-k, k]$\\

    \lstinline$x +/- m * x$ & $x_\circ = x + err_x \wedge err_x \in [-|m x|,|m x|]$\\

    \lstinline!$\sim$x! & $x_\circ$ \\

    \lstinline$!x $ & $err_x$\\

    \hline
    \lstinline! x $\diamond$ y! & $(x \diamond y) \wedge (x_\circ \diamond y_\circ)(1 + \delta_1)$ \\

    \lstinline!sqrt(x)! & $sqrt(x) \wedge sqrt(x_\circ)(1 + \delta_2)$\\

    \lstinline!val z = x! & $z = x \wedge z_\circ = x_\circ$\\

  \lstinline!if (c(x)) e1(x)! & $((c(x) \wedge e1(x)) \vee (\neg c(x) \wedge e2(x))) \wedge$ \\

  \lstinline!else e2(x)! & $((c_\circ(x_\circ) \wedge e1_\circ(x_\circ)) \vee (\neg c_\circ(x_\circ) \wedge e2_\circ(x_\circ)))$ \\

    \lstinline$g(x)$ & $g(x) \wedge g_\circ(x_\circ)$\\

    \hline

      \multicolumn{2}{c}{$\diamond \in \{+, -, *, /\}$} \\
     \multicolumn{2}{c}{$-\epsilon_m \le \delta_i \wedge \delta_i \le \epsilon_m$, all $\delta$ are fresh}\\
     \multicolumn{2}{c}{$cond_\circ$ and $e_\circ$ denote functions with roundoff errors at each step}
  \end{tabular}
  \caption{Semantics of our specification language.}
  \label{tab:semantics}
\end{table}
Table~\ref{tab:semantics} summarizes how verification constraints are generated from our specification language.
Each variable \code{x} in the specification corresponds to two real-valued variables $x, x_\circ$,
the ideal one in the absence of uncertainties and roundoff errors and the actual one,
computed by the compiled program. Note that the ideal and actual variables are related only through the 
error bounds in the pre- and postconditions, which allows for the ideal and actual executions to take
different paths.

In the method body we have to take into account roundoff errors from arithmetic operations
and the propagation of existing errors.
Our system currently supports operations $\{+, -, *, /, \sqrt{}\}$, but these can be in principle extended
to elementary functions, for instance by encoding them via Taylor expansions~\cite{Makino2003}.

Note that the resulting verification conditions are parametric in the machine epsilon.

\subsection{Specification Generation}
\label{subsec:specgen}
In order to give feedback to developers and to facilitate automatic modular analysis,
our tool also provides automatic specification generation. By this we mean that the
programmer still needs to provide the environment specification in form of preconditions,
but our tool automatically computes a precise postcondition.

Formally, we can rewrite the constraint (\ref{hornclause}) as
\begin{equation*}
\forall \vec{x}, res.\; (\exists \vec{y}.\; P(\vec{x}) \wedge body(\vec{x}, \vec{y}, res))
\to Q(\vec{x}, res)
\end{equation*}
where $Q$ is now unknown. We obtain the most precise postcondition $Q$ by
applying quantifier elimination (QE) to $P(\vec{x}) \wedge body(\vec{x}, \vec{y}, res)$ and eliminate
$\vec{y}$. The theory of arithmetic over reals admits QE so it is theoretically possible to use this approach.\\
We do not currently use a full QE procedure for specification generation, as it is expensive and it is not
clear whether the returned expressions would be of a suitable format.
Instead, we use our approximation approach which computes ranges and maximum errors in a forward fashion
and allows us to compute an (over) approximation of a postcondition of the form $res \in [a, b] \wedge res \pm u$.

\subsection{Difficulty of Simple Encoding into SMT solvers}
\label{sec:bsplines}

For small functions we can already prove interesting properties by using the exact encoding of the problem
just described and discharging the verification constraints with Z3.
Consider the following code a programmer may write to implement the third B-spline basic function which
is commonly used in signal processing~\cite{Jiang2003}.
\begin{lstlisting}[mathescape,numbers=none]
  def bspline3(u: Real): Real = {
    require(0 $\le$ u && u $\le$ 1 && u $\pm$ 1e-13)
    -u*u*u / 6.0
  } ensuring (res $\ge$ -0.17 $\le$ res && res $\le$ 0.05 && res $\pm$ 1e-11)
\end{lstlisting}
Functions and the correspoding verification conditions of this complexity are already within the possibilities
of the nonlinear solver within Z3. For more complex functions however, Z3 does not (yet) provide an answer in a reasonable time, 
or returns unknown. Whether alternative techniques in SMT solvers can help in such cases remains to be seen \cite{Jovanovic2012,
borralleras12:linear_arith_polyn_const}.
We here provide an approach based on step-wise approximation that addresses the difficulty of general-purpose constraint solving.

\subsection{Verification with Approximations}
In order to nontheless verify interesting programs, we have developed an approximation procedure that computes
a sound over-approximation of the range of an expression and of the uncertainty on the output.
This procedure is a forward computation and we also use it to generate specifications automatically.
We describe the approximation procedure in detail in Section~\ref{sec:nonlinear}, for now we will assume that it exists
and, given a precondition $P$ and an expression $expr$, computes a sound bound on the output range and its associated uncertainty:
\begin{align*}
&([a, b], err) = \text{evalWithError}(P, expr) \Leftrightarrow \\
&\forall \vec{x}, \vec{x_\circ}, res, res_\circ. P(\vec{x}, \vec{x_\circ}) \wedge res = expr(\vec{x}) \wedge
   res_\circ = expr_\circ(\vec{x_\circ})\\
     &\to \neg( res < a \vee b < res) \wedge | res - res_\circ | < err
\end{align*}

\begin{figure}[t]
  \centering
    \begin{lstlisting}[mathescape,language=pseudo,numbers=none]
    def getNextApproximation(vc, precision):
    \end{lstlisting}
    \includegraphics[width=0.47\textwidth]{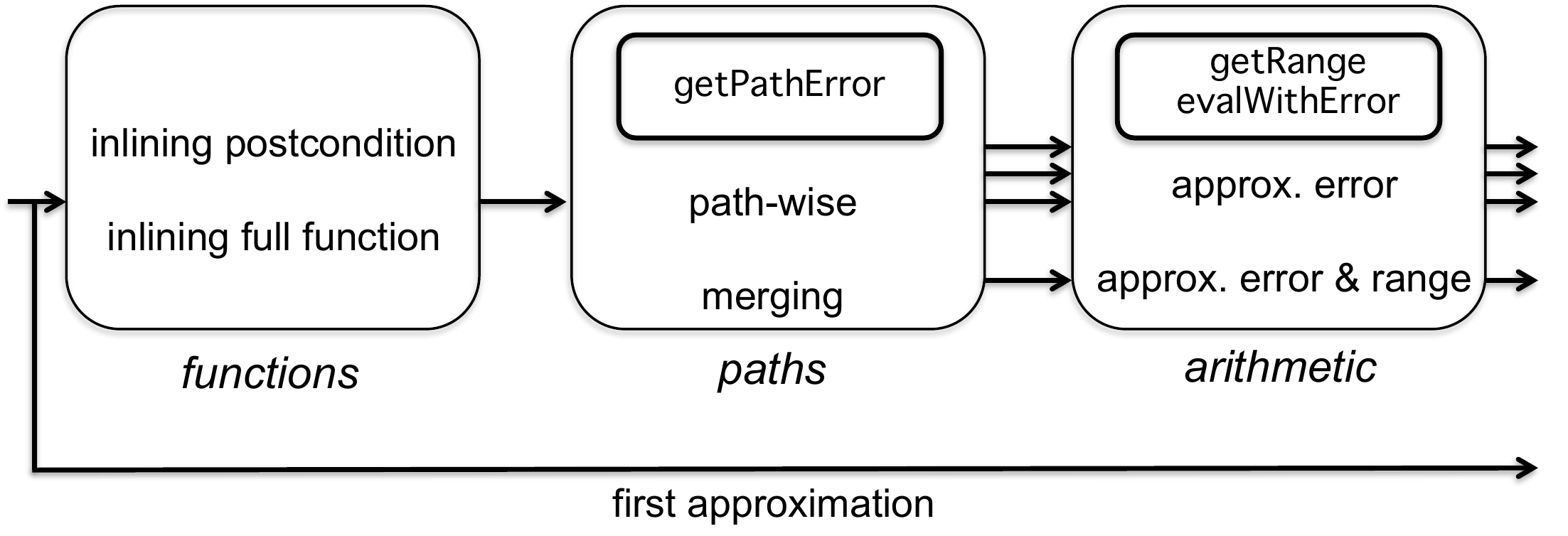}
    \caption{Approximation pipeline.}
    \label{fig:approximation-stream}
\end{figure}

We have identified
three possibiities for approximation: nonlinear arithmetic, function calls, and paths and each can be
approximated at different levels. We have observed in our experiments, that ``one size does not fit all''
and a combination of different approximations is most successful in proving the verification conditions
we encountered.
For each verification condition we thus construct approximations until Z3 is able to prove one, or until we
run out of approximations where we report the verification as failed.
We can thus view verification as a stream of approximations to be proven. We illustrate the pipeline
that computes the different approximations in Figure~\ref{fig:approximation-stream}. The routines
\code{getPathError, getRange} and \code{evalWithError} are described in the following sections in more
detail. 


The first approximation (indicated by the long arrow in Figure~\ref{fig:approximation-stream}) 
is to use Z3 alone on the entire constraint constructed by the rules in Table~\ref{tab:semantics}. This is indeed an approximation,
as all function calls are treated as uninterpreted functions in this case. As noted before, this
approach only works in very simple cases or when no uncertainties and no functions are present. 

\paragraph{Function calls}
If the verification constraint contains function calls and the first approximation failed,
our tool will attempt to inline postconditions and pass on the resulting constraint down the approximation
pipeline. We support inlining of both user-provided postconditions and postconditions computed by our own 
specification generation procedure. If this still is not precise enough, we inline the entire function body.

Postcondition inlining is implemented by replacing the function call with a fresh variable and constraining
it with the postcondition. Thus, if verification suceeds with inlining the postcondition, we avoid
having to consider each path of the inlined function separately and can perform modular verification
avoiding a potential path explosion problem. Such modular verification is not feasible when postconditions
are too imprecise and we plan to explore the generation of more precise postconditions in the future.
One step in this direction is to allow postconditions that are parametric in the initial errors, for example
with the operator $!x$ introduced in Section~\ref{sec:specification}. While our tool currently supports
postcondition inlining with such postconditions, we do not yet generate these automatically.

\paragraph{Arithmetic}
The arithmetic part of the verification constraints generated by Table~\ref{tab:semantics} can be essentially 
divided into the ideal part and the actual part, which includes roundoff errors at each computation step.
The ideal part determines whether the ideal range constraints in the postcondition are satisfied and the
actual part determines whether the uncertainty part of the postcondition is satisfied.
We can use our procedure presented in Section~\ref{sec:nonlinear} to compute a sound approximation of 
both the result's range as well as its uncertainty. Based on this, our tool constructs two approximations.
For the first, the ideal part of the constraint is left untouched and the actual part is replaced by the computed
uncertainty bound. This effectively removes a large number of variables and many times suffiently
simplifies the constraint for Z3 to succeed.
If this fails, our tool additionally replaces the ideal part by the computed range constraint.
Note that the second approximation may not have enough information to prove a more complex postconditions,
as correlation information is lost.
We note that the computation of ranges and errors is the same for both approximations and thus
trying both does not affect efficiency significantly.

\paragraph{Paths}
In the case of several paths through the program, we have the option to consider each path separately or
to merge results at each join in the control flow graph. This introduces a tradeoff between efficiency
and precision, since on one hand, considering each path separately leads to an exponential number of paths
to consider. On the other hand, merging at each join looses correlation information between variables
which may be necessary to prove certain properties. Our approximation pipeline chooses merging first,
before resorting to a path-by-path verification in case of failure. 
We believe that other techniques for exploring the path space could also be integrated into our 
tool~\cite{Kuznetsov2012,Chaudhuri2010}.
Another possible improvement are heuristics that select a different order of approximations 
depending on particular characteristics of the verification condition.



\paragraph{Example}
\begin{figure}
\begin{lstlisting}[mathescape,numbers=none]

def comparisonValid(x: Real): Real = {
  require(-2.0 < x && x < 2.0)
  val z1 = sineTaylor(x)
  val z2 = sineOrder3(x)
  z1 - z2
} ensuring(res => $\sim$res <= 0.1)

def comparisonInvalid(x: Real): Real = {
  require(-2.0 < x && x < 2.0)
  val z1 = sineTaylor(x)
  val z2 = sineOrder3(x)
  z1 - z2
} ensuring(res => $\sim$res <= 0.01) // counterexample: 1.0
  
def sineTaylor(x: Real): Real = {
  require(-2.0 < x && x < 2.0)
  x - (x*x*x)/6.0 + (x*x*x*x*x)/120.0 - (x*x*x*x*x*x*x)/5040.0 
} ensuring(res => -1.0 < res && res < 1.0 && res +/- 1e-14)

def sineOrder3(x: Real): Real = {
  require(-2.0 < x && x < 2.0)
  0.954929658551372 * x -  0.12900613773279798*(x*x*x)
} ensuring(res => -1.0 < res && res < 1.0 && res +/- 1e-14)

\end{lstlisting}
\caption{Different polynomial approximations of sine.}
\label{fig:sine-approximations}
\end{figure}

We illustrate the verification algorithm on the example in Figure~\ref{fig:sine-approximations}.
The functions \code{sineTaylor} and \code{sineOrder3} are verified first since they do not contain
function calls. Verification with the full verification constraint fails. Next, our tool computes the
errors on the output and Z3 suceeds to prove the resulting constraint with the ideal part untouched.
From this approximation our tool directly computes a new, more precise postcondition, in particular
it can narrow the resulting error to \texttt{1.63e-15}.
Next, our tool considers the \code{comparison} function. Inlining only the postcondition is not enough
in this case, but computing the error approximation on the inlined functions suceeds in verifying the
postcondition. Z3 is again able to verify the real-valued portion independently.
Finally, the tool verifies that the preconditions of the function calls are satisfied by using Z3 
alone.
Verification of the function \code{comparisonInvalid} fails with all approximations so that our tool
reports \code{unknown}. It also provides the counterexample it obtains from Z3 ($x = 1.0$), 
which in this case is a valid counterexample.

\subsection{Soundness}

 Our procedure is sound because our constraints overapproximate
 the actual errors. Furthemore, 
 even in the full constraint as generated from Table~\ref{tab:semantics},
 roundoff errors are overapproximated
 since we assume the worst-case error bound at each step. While this ensures soundness, it also
 introduces incompleteness, as we may fail to validate a specification because our overapproximation
 is too large.
 This implies that counterexamples reported by Z3 are in general only valid, if they disprove the ideal
 real-valued part of the verification constraint. In general, this is easy to distinguish from the case
 where verification fails due to too large error bounds, as Z3 prefers simpler counterexamples over complex ones and thus
 chooses all error variables to be zero if the ideal part is being violated.

\section{Solving Nonlinear Constraints}
\label{sec:nonlinear}
Having given an overview of the approximation pipeline, we now describe 
the computation of the approximation for nonlinear arithmetic, which corresponds to the last box in Figure~\ref{fig:approximation-stream}.
For completeness of presentation, we first review interval and affine arithmetic which are common choices 
for performing sound arithmetic computations and which we also use as part of our technique.
We then present our novel procedure for computing the output range of a nonlinear expression given
ranges for its inputs that can be a more precise substitute for interval or affine arithmetic.
Finally, we continue with a procedure that computes a sound overapproximation 
of the uncertainty on the result of a nonlinear expression.

One possibility to perform guaranteed computations is to use standard interval 
arithmetic~\cite{Moore1966}. Interval arithmetic computes a bounding interval for 
each basic operation as
\begin{equation*}
x \circ y = [ min(x \circ y), max( x \circ y)] \quad \quad  \circ \in \lbrace +, -, *, /  \rbrace
\end{equation*}
and analogously for square root. 
Affine arithmetic was originally introduced in~\cite{stolfi97} and addresses
the difficulty of interval arithmetic in handling correlations between
variables.
Affine arithmetic represents possible values of variables as affine forms
$$
\hat{x} = x_0 + \sum_{i=1}^n x_i \epsilon_i
$$
where $x_0$ denotes the \emph{central value} (of the represented interval) and
each \emph{noise symbol} $\epsilon_i$ is a formal variable denoting a deviation
from the central value, intended to range over $[-1, 1]$. The maximum magnitude
of each \emph{noise term} is given by the corresponding $x_i$. Note that the
sign of $x_i$ does not matter in isolation, it does, however, reflect the
relative dependence between values.  For example, take $x = x_0 + x_1
\epsilon_1$, then
\begin{align*}
x - x &= x_0 + x_1 \epsilon_1 - (x_0 + x_1 \epsilon_1) \\
&= x_0 - x_0 + x_1 \epsilon_1 - x_1 \epsilon_1 = 0
\end{align*}
If we subtracted $x = x_0 - x_1 \epsilon_1$ instead, the resulting interval
would have width $2 * x_1$ and not zero.

The range represented by an affine form is computed as
$$
[\hat{x}] = [x_0 - rad(\hat{x}), x_0 + rad(\hat{x})] ,\quad \quad  rad(\hat{x}) = \sum_{i=1}^n |x_i|
$$
A general affine operation $\alpha \hat{x}+\beta \hat{y}+\zeta$ consists of
addition, subtraction, addition of a constant ($\zeta$) or multiplication by a
constant ($\alpha, \beta$).  Expanding the affine forms $\hat{x}$ and
$\hat{y}$ we get
\begin{equation*}
 \label{eqn:affineOp}
 \alpha \hat{x}+\beta \hat{y}+\zeta=(\alpha x_0+\beta y_0+ \zeta)+\sum_{i=1}^n (\alpha x_i + \beta y_i) \epsilon_i
\end{equation*}
An additional motivation for using affine arithmetic is that different contributions
to the range it represents remain, at least partly, separated. This information can
be used for instance to help identify the major contributor of a result's uncertainty
or to separate contributions from external uncertainties from roundoff errors.

\subsection{Range Computation}
\label{subsec:ranges}
The goal of this procedure is to perform a forward-computation to determine the real-valued range of a nonlinear arithmetic expression given
ranges for its inputs. Two common possibilities are interval and affine arithmetic, but they tend to overapproximate
the resulting range, especially if the input intervals are not sufficiently small (order 1). Affine arithmetic
improves over interval arithmetic somewhat by tracking linear correlations, but in the case of nonlinear
expressions the results can become actually worse than for interval arithmetic.

\paragraph{Observation:} A nonlinear theorem prover such as the one that comes with Z3 can decide with relatively good precision whether a given
bound is sound or not. That is we can check with a prover whether for an expression $e$ the range $[a, b]$
is a sound interval enclosure. This observation is the basis of our range computation.\\

The input to our algorithm is a nonlinear expression $expr$ and a precondition $P$ on its inputs, which
specifies, among possibly other constraints, ranges on all input variables $\vec{x}$.
The output is an interval $[a, b]$ which satisfies the following:
\begin{align*}
[a, b] &= \text{getRange}(P,expr) \Leftrightarrow \\
&\forall \vec{x}, res. P(\vec{x}) \wedge res = expr(\vec{x})  \to \neg( res < a \vee b < res)
\end{align*}

\paragraph{The algorithm} for computing the lower bound of a range is given in 
Figure~\ref{alg:range-computation}. The computation for the upper bound is symmetric.
For each range to be computed, our tool first computes an initial sound estimate of the
range with interval arithmetic.
It then performs an initial quick check to test whether the computed first approximation bounds
are already tight. If not, it uses the first approximation as the starting point and then
narrows down the lower and upper bounds using a binary search.
At each step of the binary search our tool uses Z3 to confirm or reject the newly proposed bound.

The search stops when either Z3 fails, i.e. returns unknown for a query or cannot answer within
a given timeout, the difference between subsequent bounds is smaller than a precision threshold,
or the maximum number of iterations is reached. This stopping criterion can be set dynamically.

\paragraph{Additional constraints}
In addition to the input ranges, the precondition may also contain further constraints on the variables. For example consider again the
method \code{triangle} in Figure~\ref{fig:example}.
The precondition bounds the inputs as $a, b, c \in [1, 9]$, but the formula is useful only for valid triangles, i.e.
when every two sides together are longer than the third. If not, we will get an error at the latest when we try to take 
the square root of a negative number. In interval-based approaches we can only consider input intervals that satisfy this constraint for
all values, and thus have to check several (and possibly many) cases.
In our approach, since we are using Z3 to check the soundness of bounds, we can assert the additional constraints up-front
and then all subsequent checks are performed with respect to all additional and initial contraints.
This allows us to avoid interval subdivisions due to imprecisions or problem specific constraints such as those in the triangle
example. This becomes especially valuable in the presence of multiple variables, where we may need an exponential number of 
subdivisions.

\begin{figure}
\begin{lstlisting}[language=pseudo,numbers=none]
def getRange(expr, precondition, precision, maxIterations):
  z3.assertConstraint(precondition)
  [aInit, bInit] = evalInterval(expr, precondition.ranges);

  //lower bound
  if z3.checkSat(expr ¡ a + precision) == UNSAT
    a = aInit
    b = bInit
    numIterations = 0
    while (b-a) < precision $\wedge$ numIterations < maxIterations
      mid = a + (b - a) / 2
      numIterations++
      z3.checkSat(expr ¡ mid) match
        case SAT $\Rightarrow$ b = mid
        case UNSAT $\Rightarrow$ a = mid
        case Unknown $\Rightarrow$ break
    aNew = a
  else
    aNew = aInit
  
  bNew = ... //upper bound symmetrically
  return: [aNew, bNew]
\end{lstlisting}
\caption{Algorithm for computing the range of an expression.}
\label{alg:range-computation}
\end{figure}

\subsection{Error Approximation}
 \label{sec:xfloat}
We now describe our approximation procedure which, for a given expression $expr$ and
a precondition $P$ on the inputs, computes the range and error on the output. More formally,
our procedure satisfies the following:
\begin{align*}
&([a, b], err) = \text{evalWithError}(P, expr) \Leftrightarrow \\
&\forall \vec{x}, \vec{x_\circ}, res, res_\circ. P(\vec{x}, \vec{x_\circ}) \wedge res = expr(\vec{x}) \wedge
   res_\circ = expr_\circ(\vec{x_\circ})\\
     &\to \neg( res < a \vee b < res) \wedge | res - res_\circ | < err
\end{align*}
where $expr_\circ$ represents the expression evaluated in floating-point arithmetic and $\vec{x}, \vec{x_\circ}$
are the ideal and actual variables. The precondition specifies the ranges and uncertainties of initial variables
and other additional constraints on the ideal variables. The uncertainty specification is necessary, as it
related the ideal and actual variables.

The idea of our procedure is to ``execute'' a computation while keeping track of the output range of the
current expression and its associated errors. At each arithmetic operation, we propagate existing errors,
compute an upper bound on the roundoff error and add it to the overall errors. Since the roundoff error
depends proportionally on the range of values, we also need to keep track of the ranges as precisely as possible.

Our procedure is build on the abstraction that a computation is an ideal computation plus or minus some
uncertainty. The abstraction of floating-point roundoff errors that we chose also follows this separation:
\begin{align*}
  fl(x \diamond y) &= (x \diamond y)(1 + \delta) = (x \diamond y) + (x \diamond y)\delta
\end{align*}
for $\delta \in [-\epsilon_m, \epsilon_m]$ and $\diamond \in \{+, -, *, /\}$.
This allows us to treat all uncertainties in a unified manner.

Our procedure builds on the idea of the SmartFloat datatype~\cite{Darulova2011}, which uses affine arithmetic
to track both the range and the errors. For nonlinear operations, however, the so computed ranges become very pessimistic
quickly and the error computation may also suffer from this imprecision. We observed that since the errors
tend to be relatively small, this imprecision does not affect the error propagation itself to such an extent.
If the initial errors are small (less than one), multiplied nonlinear terms tend to be even smaller,
whereas if the affine terms are larger than one, the nonlinear terms grow.
We thus concentrate on improving the ideal range of values and use our novel range computation procedure
for this part and leave the error propagation with affine arithmetic as in~\cite{Darulova2011}.

In our adaptation, we represent every variable and intermediate computation result as a
  datatype with the following components:
$$
 x: (range: Interval, \hat{err}: AffineForm)
$$
where $range$ is the range of this variable, computed as described in Section~\ref{subsec:ranges}
and $\hat{err}$ is the affine form representing the errors. The (overapproximation) of the actual range including
all uncertainties is then given by $totalRange = range + [\hat{err}]$, where $\hat{err}$ denotes the interval
represented by the affine form.

\paragraph{Roundoff error computation}
Roundoff errors are computed at each computation step as
$$
\rho = \delta * maxAbs(totalRange)
$$
where $\delta$ is the machine epsilon, and added to $\hat{err}$ as a fresh noise term.
Note that this roundoff error computation makes our error computation parametric in 
the floating-point precision.

\paragraph{Error propagation}
For affine operations addition, subtraction, and multiplication by a constant factor
the propagated errors are computed term-wise and thus as for standard affine
arithmetic. We refer the reader to~\cite{stolfi97,Darulova2011} for further details and
describe here only the propagation for nonlinear arithmetic. For multiplication, division
and square root, the magnitude of errors also depends on the ranges of variables.
Since our ranges are not affine terms themselves, propagation has to be adjusted.
In the following, we denote the range of a variable $x$ by $[x]$ and its associated error by the
affine form $\hat{err_x}$. When we write $[x] * \hat{err}_y$ we mean that the interval $[x]$
is converted into an affine form and the multiplication is performed in affine arithmetic.

Multiplication is computed as
\begin{align*}
x * y &= ([x] + \hat{err_x})([y] + \hat{err_y})\\
&= [x] * [y] + [x] * \hat{err_y} + [y] * \hat{err_x} +  \hat{err_x} * \hat{err_y} + \rho
\end{align*}
where $\rho$ is the new roundoff error.
Thus the first term contributes to the ideal range and the remaining three to the error affine form.
The larger the factors $[x]$ and $[y]$ are, the larger the finally computed errors will be.
In order to keep the overapproximation as small as possible, we evaluate $[x]$ and $[y]$
with our new range computation.

Division is computed as
\begin{align*}
\frac{x}{y} &= x* \frac{1}{y} = ([x] + \hat{err_x})([1/y] + \hat{err_{1/y}})\\
&= [x] * [\frac{1}{y}] + [x] * \hat{err_{\frac{1}{y}}} + [\frac{1}{y}] * \hat{err_x} + \hat{err_x} * \hat{err_{\frac{1}{y}}} + \rho
\end{align*}
For square root, we first compute an affine approximation of square root as in~\cite{Darulova2011}
\begin{equation*}
\sqrt{x} = \alpha * x + \zeta + \theta
\end{equation*}
and then perform the affine multiplication term wise.


\paragraph{Overflows and NaN}
Our procedure allows us to detect potential overflows,
division by zero and square root of a negative value, as our tool computes ranges of all
intermediate values. We currently report these issues as warnings to the user.

\subsection{Limitations} The limitation of this approach is clearly the ability of Z3 to check our constraints.
We found its capabilities satisfactory, although we expect the performance to still significantly improve.
To emphasize the difference to the constraints that are defined by Table~\ref{tab:semantics}, the constraints
we use here do not add errors at each step and thus the number of variables is reduced significantly.
We also found several transformations helpful, such as rewriting powers (e.g. $x*x*x$ to $x^3$),
multiplying out products and avoiding non-strict comparisons in the precondition, although the benefits
were not entirely consistent.
Note that at each step of our error computation, our tool computes the current range.
Thus, even if Z3 fails to tighten the bound for some expressions, we still compute more precise bounds 
than interval arithmetic overall in most cases, as the ranges of the remaining subexpressions have already been computed more precisely.

\section{Conditional Statements}
\label{sec:branches}
In this Section we consider the difference between the ideal and actual
computation due to uncertainties on computing branch conditions and the
resulting different paths taken.
We note that the full constraint constructed according to Section~\ref{sec:specification}
automatically includes this error. Recall that the ideal and actual computations are
independent except for the initial conditions, so that it is possible that they follow
different paths through the program.

In the case of approximation, however, we compute the error on individual paths and have to
consider the error due to diverging paths separately.
If a method encodes a continous function in the usual mathematical sense then we note that
we only need to quantify errors for each path separately. Thus, if have a method~\cite{Chaudhuri2010} to
determine whether a function with conditional statements is continuous,
then our approach described so far is sufficient to provide sound error bounds.
For the case where such a procedure does not exist or fails to provide an answer, for example due 
to nonlinearity, or the function simply is not continuous, we propose the following algorithm 
to explicitly compute the difference between the ideal and the actual computation across paths.
Note that we do not assume continuity, i.e. the algorithm allows us to compute error bounds even in the
 case on non-continous functions. 

For simplicity, we present here the algorithm for the case of one conditional statement:
\begin{lstlisting}[numbers=none]
                  if (c(x) $<$ 0) f1(x)
                  else f2(x)
\end{lstlisting}
It generalizes readily to more complex expressions.
W.l.o.g. we assume that the condition is of the form $c(x) < 0$. Indeed, any conditional
of the form $c(x) == 0$ would yield different results for the ideal and actual computation for nearly any input,
so we do not allow it in our specification language.

The actual computation commits a certain error when computing the condition of the branch
and it is this error that causes some executions to follow a different branch than the
corresponding ideal one would.
Consider the case where the ideal computation evaluates f1, but the actual one evaluates f2.
Algorithm~\ref{alg:patherror} gives the computation of the {\em path error} in this case. 
The idea is to compute the ranges of f1 and f2, but only for the inputs that could be 
diverging. The final error is then the maximum difference of these value.
The algorithm extends naturally to several variables. In the case of several paths through the
program, this error has to be, in principle, computed for each combination of paths.
We use Z3 to rule out infeasible paths up front so that the path error computation is only
performed for those paths that are actually feasible. 

\begin{figure}
\begin{lstlisting}[mathescape, language=pseudo]
def getPathError:
Input: pre ($x \in [a, b] \wedge x \pm n$)
       program (if (cond(x) $<$ 0) f1(x) else f2(x))
  val pathError1 = computePathError(pre, cond, f1, f2)
  val pathError2 = computePathError(pre, $\neg$ cond, f2, f1)
  return max (pathError1, pathError2)

def computePathError(pre, c, f1, f2):
  ([c], $err_{c}$) = evalWithError(pre, c)
  ([f2]$_{float}$, $err_{float}$) =
    evalWithError(pre $\wedge$ c(x) $\in$ [0, $err_{c}$], f2)
  [f1]$_{real}$ = 
    getRange(pre $\wedge$ c(x) $\in$ [$- err_{c}$, 0], f1)
  return: max $|$[f1]$_{real}$ - ([f2]$_{float}$ + $err_{float})|$
\end{lstlisting}
\caption{Computing error due to diverging paths.}
\label{alg:patherror}
\end{figure}

We have currently implemented this approach in our tool for the case when we use merging to
handle paths in order to avoid having to consider an exponential number of path combinations.
We also use a higher default precision and number of iterations threshold during the binary search
in the range computation as this computation requires in general very tight intervals for each
path.

We identify two challenges for performing this computation:
\begin{enumerate}
\item
 As soon as the program has multiple variables, the inputs for the different branches are
 not two-dimensional intervals anymore, which makes an accurate evaluation of the individual
 paths difficult in standard interval arithmetic. 

\item 
 The inputs for the two branches are inter-dependent. Thus, simply evaluating
 the two branches with inputs that are in the correct ranges, but are not correlated, 
 yields pessimistic results when computing the final difference (line 16).
\end{enumerate}

We overcome the first challenge with our range computation which takes into
account additional constraints. For the second challenge, we use our range computation as well,
however unfortunately Z3 fails to tighten the final range to a satisfactory precision due to timeouts. 
We still obtain much better error estimates
than with interval arithmetic alone, as the ranges of values for the individual paths
are already computed much more precisely. We report in Section~\ref{sec:experiments}
on the type of programs whose verification is already in our reach today.

\section{Experiments}
\label{sec:experiments}
The examples in Figure~\ref{fig:example} and ~\ref{fig:sine-approximations} and Section~\ref{sec:bsplines} provide an idea of 
the type of programs our tool is currently able to verify fully automatically. The B-spline example from Section~\ref{sec:bsplines}
is the largest meaningful example we were able to find that Z3 alone could verify in the presence of uncertainties.
For all other cases, it was necessary to use our approximation methods.

\subsection{Evaluating Effectiveness on Nonlinear Expressions}

To evaluate our range and error computation technique we have chosen several nonlinear expressions commonly used 
in physics, biology and chemistry~\cite{Woodford2012, Quarteroni2010, Murray2002} as benchmark functions, as well as benchmarks 
used in control systems~\cite{Anta2010} and suitable benchmarks from~\cite{DSilva2012}.
Experiments were performed on a desktop computer running Ubuntu 12.04.1 with a 3.5GHz i7 processor and 16GB of RAM.
Running times highly depend on the timeout used for Z3. Our default setting is 1 second; we did not find much improvement in the success rate above this threshold.

\begin{table*}
\centering
  \begin{tabular}{l c c c c c}
  Benchmark & Our range & interval arithmetic & Simulated range \\
\hline
doppler1  & [-137.639, -0.033951]   & [-158.720, -0.029442] &  [-136.346, -0.035273] \\

doppler2  & [-230.991, -0.022729]  & [-276.077, -0.019017] & [-227.841,-0.023235]\\

doppler3 & [-83.066, -0.50744]  & [-96.295, -0.43773] & [-82.624, -0.51570] \\

rigidBody1 & [-705.0, 705.0] & [-705.0, 705.0] & [-697.132, 694.508]\\

rigidBody2 & [-56010.1, 58740.0] & [-58740.0, 58740.0] & [-54997.635, 57938.052]\\

jetEngine & [-1987.022, 5099.243]  & $[- \infty, \infty]$ & [-1779.551, 4813.564]\\

turbine1  & [-18.526, -1.9916]  & [-58.330, -1.5505] & [-18.284, -1.9946]\\

turbine2  & [-28.555, 3.8223]   & [-29.437, 80.993] & [-28.528, 3.8107]\\

turbine3 & [0.57172, 11.428] & [0.46610, 40.376] & [0.61170, 11.380]\\

verhulst  & [0.31489, 1.1009]   & [0.31489, 1.1009] & [0.36685,0.94492]\\

predatorPrey & [0.039677, 0.33550]     &[0.037277, 0.35711] & [0.039669,0.33558]\\

carbonGas & [4.3032 e6, 1.6740 e7]  & [2.0974 e6, 3.4344 e7] & [4.1508 e6, 1.69074 e7]\\

Sine & [-1.0093, 1.0093]   & [-2.3012, 2.3012] & [-1.0093, 1.0093]\\

Sqrt & [1.0, 1.3985]  & [0.83593, 1.5625] & [1.0, 1.3985]\\

Sine (order 3 approx.) & [-1.0001, 1.0001]  & [-2.9420, 2.9420]  & [-1.0, 1.0] \\ 

  \end{tabular}
  \caption{Comparison of ranges computed with out procedure against interval arithmetic and simulation.
  Simulations were performed with $10^7$ random inputs. Ranges are rounded outwards.}
  \label{tab:ranges}
\end{table*}

\begin{table}
\centering
  \begin{tabular}{l l c}
  Benchmark & Our error (IA only) & Simulated error \\
  \hline
doppler1*  & 2.36e-6 &  5.97e-7\\
doppler1   &  4.92e-13 (4.95e-13) &  7.11e-14\\
doppler2*  &  6.21e-5 &   1.85-5\\
doppler2  &  1.29e-12 &   1.14e-13\\
doppler3* &  1.23e-4 &   5.96e-5\\
doppler3 &  2.03e-13 (2.05e-13) &   4.27e-14\\
rigidBody1* &  9.21e-7 &  8.24e-7  \\
rigidBody1 &  5.08e-13 &  2.28e-13 \\
rigidBody2*  &  1.51e-4 &   1.25e-4 \\
rigidBody2 &  6.48e-11 &   2.19e-11 \\
jetEngine*  &  0.15 \hspace{20pt}(-) &   3.58e-5\\
jetEngine  &  1.62e-8 \hspace{10pt}(-) &   5.46e-12 \\
turbine1*  &   4.86e-6 & 3.71e-7   \\
turbine1 & 1.25e-13 (1.38e-13) &  1.07e-14 \\
turbine2*  &   8.05e-6 &  7.66e-7  \\
turbine2 &  1.76e-13 (1.96e-13) & 1.43e-14  \\
turbine3*  &  3.35e-6 &  1.04e-6  \\
turbine3 &  8.50e-14 (9.47e-14) &  5.33e-15  \\
verhulst*  &  2.82e-4 &  2.40e-4 \\
verhulst  &  6.82e-16 &  2.23e-16 \\
predatorPrey* &  9.22e-5 &  8.61e-5 \\
predatorPrey & 2.94e-16 (2.96e-16) & 1.12e-16 \\
carbonGas*  &     2114297.84 &   168874.70\\
carbonGas  &  4.64e-8 (5.04e-8) &  3.73e-9\\
Sine (single) &  1.03e-6 (1.57e-6) &   1.79e-7 \\
Sine  &         9.57e-16 (1.46e-15) &   4.45e-16 \\
Sqrt (single) &  9.03e-7 (9.52e-7) &  2.45e-7  \\
Sqrt  &  8.41e-16 (8.87e-16) &  4.45e-16  \\
Sine, order 3 (single) & 1.19e-6 (1.55e-6) & 2.12e-7\\
Sine, order 3 & 1.11e-15 (1.44e-15) & 3.34e-16\\

 \end{tabular}
  \caption{Comparison of errors computed with our procedure against simulated errors.
  Simulations were performed with $10^7$ random inputs. (*) indicates that inputs have
  external uncertainties associated.}
  \label{tab:errors}
\end{table}

\paragraph{Range computation} Stepwise estimation of errors crucially depends on the
estimate of the ranges of variables. The strength of using a constraint
solver such as Z3 is that it can perform such estimation while taking into account
the precise dependencies between variables in preconditions and path conditions.
Table~\ref{tab:ranges} compares results of our range computation procedure described in Section~\ref{sec:nonlinear}
against ranges obtained with standard interval arithmetic. Interval arithmetic is one
of the methods used for step-wise range estimation; an alternative being
affine arithmetic.
We have also experimented with an affine arithmetic implementation~\cite{Darulova2011}. 
However, we found
that affine arithmetic gives more pessimistic results for computing
ranges for non-linear benchmarks. 
We believe that this is due to imprecision in computing nonlinear operations. Note,
however, that we still use affine arithmetic to estimate errors given the computed
ranges.

We set the default precision threshold to \code{1e-10} and maximum
number of iterations for the binary search to 50. To obtain an 
idea about the ranges
of our functions, we have also computed a lower bound on the range
using simulations with $10^7$ random inputs and with exact rational arithmetic evaluation
of expressions.
We observe that our range computation can significantly improve over standard interval bounds. 
The \code{jetEngine}
benchmark is a notable example, where interval arithmetic yields the bound $[- \infty, \infty]$, but our procedure
can still provide bounds that are quite close to the true range.
Running times are below 7 seconds for the most complex benchmarks, except for \code{jetEngine} which runs in about
1 minute due to timeouts from Z3 for some intermediate ranges.

\paragraph{Error computation}
Table~\ref{tab:errors} compares uncertainties computed by our tool against maximum uncertainties obtained
through extensive simulation with $10^7$ random inputs. We ran the simulation in parallel with rational and
their corresponding floating-point value and obtained the error by taking the difference in the result.
Benchmarks marked with (*) have added initial uncertainties.
Unless otherwise indicated, we used 
double floating-point precision. To our knowledge this is the first
quantitative comparison of an error computation precision with (an approximation) of the 
true errors on such benchmarks.
Except for the benchmarks \code{jetEngine*} our computed uncertainties are within an order
and many times even closer to the underapproximation of the true errors provided by simulation.
In the case of the \code{jetEngine*} benchmark, we believe that the imprecision is mainly due to its complexity
and subsequent failures of Z3.
The values in parentheses in the second column indicate errors computed if ranges at each arithmetic operation
are computed using interval arithmetic alone. While we have not attempted to improve the affine arithmetic-based 
error computation from~\cite{Darulova2011}, we can see that in some cases a more precise range computation can gain 
us improvements. The full effect of the imprecision of standard range computation appears 
when, due to this imprecision,
we obtain possible errors such as division-by-zero or 
square root of a negative number errors. The 
first case happens in the case of the non-linear
\code{jetEngine} benchmark, so with interval arithmetic alone we would therefore not 
obtain any meaningful result.
Similarly, for the triangle example from Section~\ref{sec:example}, without being able to 
constrain the inputs
to form valid triangles, we cannot compute any error bound, because the radicand 
becomes possibly negative.

\renewcommand{\arraystretch}{1.3}
\begin{table}
\centering
  \begin{tabular}{c c c }
  Benchmark & Range & Max. abs. error\\
  \hline
triangle1 (0.1) & [0.29432, 35.0741]      & 2.72e-11 \\
triangle2 (1e-2) & [0.099375, 35.0741]      & 8.04e-11 \\
triangle3 (1e-3) & [3.16031e-2, 35.0741]    & 2.53e-10 \\
triangle4 (1e-4) & [9.9993e-3, 35.0741]     & 7.99e-10 \\
triangle5 (1e-5) & [3.1622e-3, 35.0741]     & 2.53e-9 \\
triangle6 (1e-6) & [9.9988e-4, 35.0741]     & 7.99e-9 \\
triangle7 (1e-7) & [3.1567e-4, 35.0741]     & 2.54e-8 \\
triangle8 (1e-8) & [9.8888e-5, 35.0741]     & 8.08e-8 \\  
triangle9 (1e-9) & [3.0517e-5, 35.0741]     & 2.62e-7 \\
triangle10 (1e-10) & - & - \\
  \end{tabular}
  \caption{Ranges and errors for increasingly flat triangles. All values are rounded outwards.
  Interval arithmetic alone fails to provide any result.}
  \label{tab:triangleprogression}
\end{table}
Table~\ref{tab:triangleprogression} presents another relevant experiment, 
evaluating the ability to use additional
constraints during our range computation. We use the triangle example
from Section~\ref{sec:example} with additional constraints allowing increasingly flat triangles by setting
the threshold on line 13 (\code{a + b > c + 1e-6}) to the different values given in the first column.
As the triangles become flatter, we observe an expected increase in uncertainty on the input since the formula 
becomes more prone to roundoff errors. At threshold \code{1e-10} our range computation fails to
provide the necessary precision and the radicand becomes possibly negative.
(We used double precision in this example as well.)

\subsection{Evaluating Errors across Program Paths}

\begin{figure}
\begin{lstlisting}[mathescape]
def cav10(x: Real): Real = {
  require(x.in(0, 10))
  if (x*x - x >= 0) 
    x/10
  else 
    x*x + 2
} ensuring(res => 0 <= res && res <= 3.0 && res +/- 3.0)
  
def squareRoot3(x: Real): Real = {
  require( x.in(0,10) && x +/- 1e-10 )
  if (x < 1e-5)
    1 + 0.5 * x
  else 
    sqrt(1 + x)
} ensuring( res => res +/- 1e-10) 

def smartRoot(a: Real, b: Real, c: Real): Real = {
  require(3 <= a && a <= 3 && 3.5 <= b && b <= 3.5 &&
   c.in(-2, 2) && b*b - a * c * 4.0 > 0.1)

  val discr = b*b - a * c * 4.0
  if(b*b - a*c > 10.0) {
    if(b > 0.0) c * 2.0 /(-b - sqrt(discr))
    else if(b < 0.0)  (-b + sqrt(discr))/(a * 2.0)
    else (-b + sqrt(discr))/(a * 2.0)
  }
  else {
    (-b + sqrt(discr))/(a * 2.0)
  }
} ensuring (res => res +/- 6e-15)
\end{lstlisting}
\caption{Path error computation examples.}
\label{fig:robustness-examples}
\vspace{-20pt}
\end{figure}
Figure~\ref{fig:robustness-examples} presents several examples to evaluate our error computation procedure
across different paths from Section~\ref{sec:branches}.
The first method \code{cav10}~\cite{Ghorbal2010} has been used before as a benchmark function for computing the
output range. Our tool can verify the given postcondition immediately. Note that the error on the result
is actually as large as the result itself, since the method is non-continuous, an aspect that has been ignored
in previous work, but that our tool detects automatically.
The method \code{squareRoot3} is also an non-continous function that computes the square root of $1 + x$
using an approximation for small values and the regular library method otherwise. Note the additional uncertainty
on the input, which could occur for instance if this method is used in an embedded controller. Our tool can verify the
given spefication. If we change the condition on line 10 to \code{x < 1e-4} however, verification fails.
In this fashion, we can use our tool to determine the appropriate branch condition to meet the precision requirement.
The above examples verify all in under 5 seconds.
Finally, the \code{smartRoot} method computes one root of a quadratic equation using the well-known more precise method
from \cite{Goldberg1991}. We are currently not aware of an automatic tool to prove programs continuous in the
presence of nonlinear arithmetic, so that we need to compute the error across different paths as well.
Our tool succeeds in verifying the postcondition in about 25s. The rather long running time is due to the complexity of the
conditions when computing the error across paths, and thus Z3's longer response time, and a number of Z3 timeouts (Z3 timeout
here means merely that some ranges have not been tightend to the best precision). 
In the future, we envision that optimizations that select alternative approximations 
can be performed automatically by a trustworthy compiler.

\section{Related work}
\label{sec:related}
Current approaches for verifying floating-point code include abstract interpretation,
interactive theorem proving and decision procedures, which we survey in this section.
We are not aware of work that would automatically integrate reasoning about
uncertainties.

\paragraph{Abstract interpretation (AI)}
Abstract domains that are sound with respect to floating-point computations can prove bounds on the
ranges of variables~\cite{Blanchet2003, Chen2009, Jeannet2009, Mine2004, Feret2004}.
The only work in this area that can also quantify roundoff errors is the tool
Fluctuat\cite{Delmas2009,Goubault2012}. These techniques use interval or affine arithmetic and together
with the required join and meet operations may yield too pessimistic results.
\cite{Ponsini2012} improves the precision of Fluctuat by refining the input domains
with a constraint solver. Our approach can be viewed as approaching the problem from a different end,
starting with an exact constraint and then using approximation until the solver succeeds.
Unlike AI tools in general, our system currently handles only functional code, in particular it does
not handle loops. If the user can provide inductive postconditions, then we can still prove the code correct,
but we do not in general discover these ourselves. Our focus lies on proving precise
bounds on the ranges in the presence of nonlinear computations and the quantification of roundoff
errors and other uncertainties.

\paragraph{Theorem proving}
The Gappa tool~\cite{Linderman2010, Dinechin2011} generates a proof checkable by
the interactive theorem prover Coq from source code with specifications.
It can reason about properties that can be reduced to
reasoning about ranges and errors, but targets, very precise
properties of specialized functions, such as software implementations of elementary
functions. The specification itself requires expertize and the proofs human
intervention. A similar approach is taken by \cite{Ayad2010} which generate verification
conditions that are discharged by various theorem provers.
Harisson has also done significant work on proving floating-point programs in the HOL Light
theorem prover~\cite{Harrison2006}.

Our approach makes a different compromise on the precision vs. automation tradeoff,
by being less precise, but automatic. The Gappa approach can be used complementary
to ours, in that if we detect that more precision is needed, Gappa is employed by
an expert user on selected methods, and the results are then used by our tool
instead of automatically computed specifications.

\paragraph{Range computation}
The Gappa tool and most constraint solvers internally use interval arithmetic for
sound range computations, whose limitations are well-known.
~\cite{Duracz2008} describes an arithmetic based on function enclosures
and~\cite{Makino2003} use an arithmetic based on taylor series as an alternative.
This approach is useful when checking a constraint, but is not suitable for a
{\em forward computation} of ranges and errors.

\paragraph{Decision procedures}
An alternative approach to verification via range computation are floating-point
decision procedures. Bit-precise constraints, however, become
very large quickly. ~\cite{Brillout2009} addresses this problem by using a combination
of over- and underapproximations.
~\cite{Haller2012} present an alternative approach in combining interval constraint solving with
a CDCL algorithm and ~\cite{Gao2010} is a decision procedure for nonlinear real arithmetic
combining interval constraint solving with an SMT solver for linear
arithmetic.\cite{Rummer2010} formalizes the floating-points for the SMT-LIB format.
While these approach can check ranges on numeric variables, they do not handle roundoff errors or
other uncertainties and cannot compute specifications automatically.

Our techniques rely on the performance of Z3. We hope that an integration of the recent 
new improved solver for nonlinear arithmetic~\cite{Jovanovic2012} will make many more verification 
problems feasible with our techniques. An alternative to this approach
is using linear approximations to solve polynomial constraints \cite{borralleras12:linear_arith_polyn_const}.
We believe that such advances are largely orthogonal to our use of range arithmetic and complement
each other.

\paragraph{Testing}
Symbolic execution is a well-known technique for generating test inputs.
\cite{Borges2012} use a combination of meta-heuristic search and interval constraint
solving to solve the floating-point constraints that arise, whereas~\cite{Lakhotia2010}
combine random search and evolutionary techniques.
\cite{Tang2010} test numerical code for precision by perturbing low-order bits of values
and rewriting expressions. The idea is to exagerate initial errors and thus make imprecisions
more visible. Probabilistic arithmetic~\cite{Scott2007} is a similar approach but it
does the perturbation by using different rounding modes.
\cite{Benz2012} also propose a testing produce to detect accuracy problems by instrumentring
code to perform a higher-precision computation side by side with the regular computations.
While these approaches are sound with respect to floating-point arithmetic, they only
generate or can check individual inputs and are thus not able to verify or compute output ranges
or their roundoff errors.

\paragraph{Robustness analysis}
~\cite{Ivancic2010} combines abstract interpretation with model checking to check
programs for stability by tracking the evolution of the width of the interval representing
a single input. \cite{Majumdar2010} use concolic execution to find inputs which, given maximum
deviations on inputs, maximize the deviation on the outputs. These two works however, use
a testing approach and cannot provide sound guarantees.
\cite{Chaudhuri2010} presents a framework for continuity analysis of programs along the 
mathematical $\epsilon-\delta$ definition of continuity and \cite{Chaudhuri2011} builds on this
work and presents a sound robustness analysis. This framework provides a syntactic proof of 
robustness for programs over reals and thus does not consider floating-points.
Our approach describes a quantitative measure of robustness for nonlinear programs with floating-point 
numbers and other uncertainties, and we believe that it can complement the cited framework.

\section{Conclusion}

We have presented a programming model for numerical
programs that decouples the mathematical problem
description from its realization in finite precision. The model
uses a \code{Real} data type that corresponds to mathematical real numbers.
The developer specifies the program using reals and indicates the target precision;
the compiler chooses a floating point representation while
checking that the desired precision targets are met. We have
described the soundness criteria by translating programs
with precision requirements into verification conditions
over mathematical reals. The resulting
verification conditions, while a natural description of the
problem being solved, are difficult to solve using a
state-of-the art SMT solver Z3. We therefore developed an
algorithm that combines SMT solving with range computation.
Our notion of soundness incorporates full input/output
behavior of functions, taking into account that, due to
conditionals, small differences in values can lead to
different paths being taken in the program. For such cases
our approach estimates a sound upper bound on the total
error of the computation.

We have evaluated our techniques on a number of benchmarks
from the literature, including benchmarks from physics,
biology, chemistry, and control systems. We have found that
invocation of SMT solver alone is not sufficient to handle
these benchmarks due to scalability issues, whereas the use
of range arithmetic by itself is not precise enough. By
combining these two techniques we were able to show that a
floating point version of the code conforms to the
real-valued version with reasonable precision requirements.

We believe that our results indicate that it is reasonable
to introduce \code{Real}s as a data type, following a list of
previously introduced mathematical abstractions in
programming languages, such as unbounded integers,
rationals, and algebraic data types. The feasibility of
verified compilation of our benchmarks suggests that it is
realistic to decouple the verification of executable
mathematical models over reals from their sound
compilation. We therefore expect that this methodology will
help advance rigorous formal verification of numerical
software and enable us to focus more on high-level
correctness properties as opposed to run-time errors alone.

\bibliographystyle{abbrvnat}
\bibliography{main}

\end{document}